\documentclass[prl,superscriptaddress,amsfonts,twocolumn,showpacs,floatfix]{revtex4}

\usepackage{graphicx,color}
\usepackage{amsmath}
\usepackage{amssymb}
\usepackage{bm}
\usepackage{ulem}

\usepackage{txfonts}

\def\Vec#1{\bm{#1}}
\def\Hc2{H_\mathrm{c2}}
\def\Tc{T_\mathrm{c}}

\begin{document}

\title{Multiband Superconductivity with Unexpected Deficiency \\of Nodal Quasiparticles in CeCu$_2$Si$_2$}

\author{Shunichiro Kittaka}
\affiliation{Institute for Solid State Physics (ISSP), University of Tokyo, Kashiwa, Chiba 277-8581, Japan}
\author{Yuya Aoki}
\affiliation{Institute for Solid State Physics (ISSP), University of Tokyo, Kashiwa, Chiba 277-8581, Japan}
\author{Yasuyuki Shimura}
\affiliation{Institute for Solid State Physics (ISSP), University of Tokyo, Kashiwa, Chiba 277-8581, Japan}
\author{Toshiro Sakakibara}
\affiliation{Institute for Solid State Physics (ISSP), University of Tokyo, Kashiwa, Chiba 277-8581, Japan}
\author{Silvia~Seiro}
\affiliation{Max Planck Institute for Chemical Physics of Solids, 01187 Dresden, Germany}
\author{Christoph Geibel}
\affiliation{Max Planck Institute for Chemical Physics of Solids, 01187 Dresden, Germany}
\author{Frank Steglich}
\affiliation{Max Planck Institute for Chemical Physics of Solids, 01187 Dresden, Germany}
\author{Hiroaki Ikeda}
\affiliation{Department of Physics, Kyoto University, Kyoto 606-8502, Japan}
\author{Kazushige Machida}
\affiliation{Department of Physics, Okayama University, Okayama 700-8530, Japan}

\date{\today}

\begin{abstract}
Superconductivity in the heavy-fermion compound CeCu$_2$Si$_2$ is a prototypical example of Cooper pairs formed by strongly correlated electrons. 
For more than 30 years, it has been believed to arise from nodal $d$-wave pairing mediated by a magnetic glue. 
Here, we report a detailed study of the specific heat and magnetization at low temperatures for a high-quality single crystal. 
Unexpectedly, the specific-heat measurements exhibit exponential decay with a two-gap feature in its temperature dependence, along with a linear dependence as a function of magnetic field and the absence of oscillations in the field angle, 
reminiscent of multiband full-gap superconductivity.
In addition, we find anomalous behavior at high fields, attributed to a strong Pauli paramagnetic effect. 
A low quasiparticle density of states at low energies with a multiband Fermi-surface topology would open a new door into electron pairing in CeCu$_2$Si$_2$. 
\end{abstract}

\pacs{74.70.Tx, 74.25.Bt, 74.25.Op}

\maketitle

%################### Introduction #####################

After the first discovery of heavy-fermion superconductivity in CeCu$_2$Si$_2$~\cite{Steglich1979PRL}, 
a number of unconventional superconductors, such as high-$\Tc$ cuprates, iron-pnictides, organic, and heavy-fermion superconductors, have been found.
Among the various issues on these novel superconductors, the identification of the superconducting gap structure is one of the most important subjects 
because it is closely related to the pairing mechanism.
Particularly, the gap symmetry of CeCu$_2$Si$_2$ has attracted attention 
because superconductivity in this compound emerges near an antiferromagnetic (AFM) quantum critical point
and heavy quasiparticles (QPs) couple to quantum critical spin excitations~\cite{Stockert2011NatPhy}. 

Up to now, the gap symmetry of CeCu$_2$Si$_2$ was inferred to be an even-parity $d$-wave type with line nodes. 
The well resolved decrease in the NMR Knight shift below the transition temperature $\Tc\simeq 0.6$~K~\cite{Ueda1987JPSJ}
is a strong evidence for the spin part of the Cooper pairs being a singlet. 
Indeed, the low-$T$ saturation of the upper critical field $\Hc2$ is attributed to the Pauli paramagnetic effect due to the spin-singlet pairing.
Based on the $T^3$ dependence of the nuclear relaxation rate $1/T_1$ and the absence of a coherence peak~\cite{Kitaoka1986JPSJ,Ishida1999PRL,Fujiwara2008JPSJ}, 
the superconducting gap was proposed to possess line nodes. 
Presently the debate is whether the gap symmetry is $d_{x^2-y^2}$ or $d_{xy}$ type~\cite{Vieyra2011PRL, Eremin2008PRL}.
However, the presence of line nodes as well as the symmetry of the gap has not yet been studied precisely using low-$T$ thermodynamic properties.

%################### Results #####################
%%%%%%% absence of nodes %%%%%%%

To elucidate the gap structure of CeCu$_2$Si$_2$, 
the specific heat $C$ in magnetic fields $H$ is herein measured at temperatures down to 40~mK using a high-quality single-crystalline sample.
Measurement of $C$ probes the QP density of states (DOS) that depends on the nodal structure.
An \textit{S}-type single crystal (having a mass of 13.8~mg) was used that presents only a superconducting ground state without magnetic ordering, 
since other types of CeCu$_2$Si$_2$ show additional contributions in $C(T)$ at low temperatures that make the interpretation of the data difficult. 
Growth and characterization of the single crystal is described in Ref.~\onlinecite{Seiro2010PSSB}.
The specific heat was measured by the standard quasi-adiabatic heat-pulse method.
The dc magnetization was measured using a high-resolution capacitive Faraday magnetometer with a vertical field gradient of 5~T/m.
All the measurements were done at ISSP.

Figure 1(a) plots the $T$ dependence of the nuclear-subtracted specific heat $C_{\rm e}=C-C_{\rm n}$ divided by $T$, 
measured with various fields applied along the [100] axis [as explained in Supplemental Material (I)].
In the normal state, $C_{\rm e}/T$ gradually increases upon cooling.
This non-Fermi-liquid behavior arises from three-dimensional spin-density-wave fluctuations occurring in the vicinity of an AFM quantum critical point~\cite{Gegenwart1998,Arndt2011PRL}. 

\begin{figure}
\includegraphics[width=3.1in]{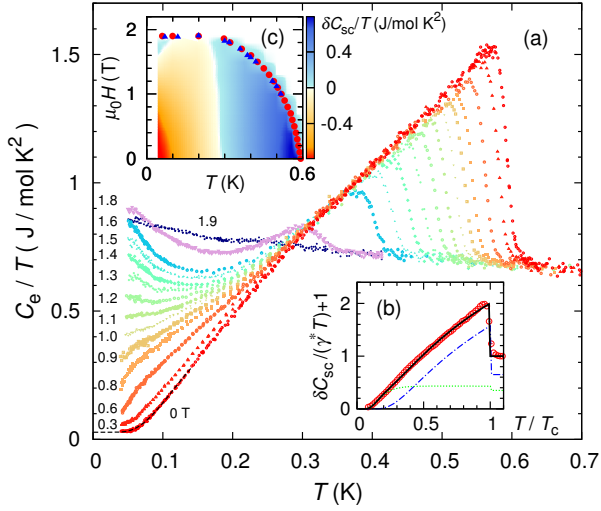}
\caption{
(Color online) 
(a)~Electronic specific heat of an \textit{S}-type CeCu$_2$Si$_2$ single crystal divided by temperature, $C_{\rm e}/T$, as a function of $T$ measured in $H \parallel [100]$. 
The dashed line is a fit to the low-$T$ part of the $C_{\rm e}(T)$ data at 0~T using the BCS formula $C_{\rm e}(T)=A\exp(-\Delta_0/T)+\gamma_0T$.
(b) Temperature variation of $\delta C_{\rm sc} (T, H)/\gamma^\ast T+1$ at $H=0$ and a best fit to the two-gap BCS model (solid line). 
Here, $\delta C_{\rm sc}(T, H)=C_{\rm e}(T, H) - C_{\rm e}(T, 3\ {\rm T})$ and $\gamma^\ast=0.84$~J/(mol K$^2$) 
are introduced to satisfy entropy balance in the BCS framework. 
The gradual increase of the normal-state $C_{\rm e}/T$ upon cooling is included in $\delta C_{\rm sc}(T, H)$.
The contribution of each gap to the total specific heat is also shown.
(c)~Field-temperature phase diagram for $H \parallel [100]$ determined by the specific-heat (circles) and magnetization (triangles) measurements.
A contour plot of $\delta C_{\rm sc}(T,H)/T$ in the superconducting state is shown using the data from (a).
Anomalous $\delta C_{\rm sc} > 0$ behavior that can be ascribed to the Pauli paramagnetic effect is clearly seen in the high-$H$ and low-$T$ region below 0.1~K.
}
\label{sh}
\end{figure}

Consider the zero-field data.
Although $\Tc=0.6$~K is slightly lower than the optimum value for this compound ($\sim0.65$~K), 
the sample shows a lower residual DOS at the base temperature and a sharper transition at $\Tc$ than those of previous reports~\cite{Bredl1983JMMM,Arndt2011PRL}. 
These facts indicate the high quality of the present sample with no significant impurities.
The specific-heat jump at $\Tc$ is found to be $\Delta C_{\rm e}(\Tc)/\gamma\Tc \sim 1.2$, slightly smaller than the weak-coupling BCS prediction of 1.43. 
At intermediate temperatures, $C_{\rm e}/T$ exhibits a nearly linear $T$ dependence 
that is consistent with the $T^3$ dependence of $1/T_1$ observed down to 0.1~K~\cite{Kitaoka1986JPSJ,Ishida1999PRL,Fujiwara2008JPSJ}.

At lower temperatures, however, $C_{\rm e}/T$ shows a large positive curvature, in contrast to the linear behavior predicted for a line-node gap [see Supplemental Material (II)].
The data can be fit using the BCS function $C_{\rm e}=A\exp(-\Delta_0/T)+\gamma_0T$ with $\Delta_0=0.39$~K and $\gamma_0=0.028$~J/(mol~K$^2$) [dashed line in Fig.~1(a)]. 
Comparison with previous results~\cite{Arndt2011PRL} shows that this positive curvature is insensitive to sample quality, i.e., to a change in $\gamma_0$ in the range 0.01 to 0.08~J/(mol~K$^2$),
which would originate from non-superconducting inclusions in the sample. 
Therefore, its origin cannot be attributed to the impurity-scattering effect of line-node superconductors~\cite{Kubert1998SSC}. 
Furthermore, extrapolating the linear behavior in $C(T)/T$ versus $T$ observed in the range $80~{\rm mK} \le T \le 250~~{\rm mK}$ to $T = 0$ results in a negative intercept, 
not only for the present data, but for all published \textit{S}-type samples (e.g., Ref.~\onlinecite{Arndt2011PRL}). 
This implies the crossover to a high power law below 80~mK, proving the intrinsic QP DOS at low energy to be extremely small in CeCu$_2$Si$_2$.

On the basis of a phenomenological two-gap model within the conventional BCS framework~\cite{Bouquet2001EPL}, 
the $T$ dependence of $C_{\rm e}/T$ including the linear behavior in the intermediate-$T$ region can be reproduced [Fig.~1(b)] 
using two BCS gaps, $\Delta_1/k_{\rm B}\Tc=1.76$ and $\Delta_2/k_{\rm B}\Tc=0.7$, whose weights are 65\% and 35\% of the total DOS, respectively. 
A signature of multiband superconductivity is found in the dependence of $C_{\rm e}/T$ with $H$ as well: 
$C_{\rm e}/T$ at 0.6~T shows a kink at 65~mK and decreases rapidly with cooling.
The $T$ variation of $C_{\rm e}/T$ matches with the prediction of two-gap superconductivity in the absence of nodal QPs.
Furthermore, the $T^3$-like dependence of $1/T_1$ down to $\sim 0.2\Tc$ is naturally led by the multiband full-gap model, 
as is demonstrated for the iron-pnictide superconductors~\cite{Yashima2009JPSJ};
$1/T_1$ measurement below 100~mK is desired to further confirm a fully-opened multigap.

Direct evidence for the deficiency of nodal QP excitations comes from the field variation of $C_{\rm e}(H)$ at 60~mK. 
For both field orientations parallel and perpendicular to the [001] axis, 
$C_{\rm e}$ gradually increases in proportion to $H$ for fields up to about 0.2~T (Fig.~2),
in contrast to the $ \sqrt{H}$ variation of $C_{\rm e}(H)$ for nodal superconductors~\cite{Volovik1993JETPL}.
The upward curvature at 0.5~T is ascribed to the abrupt enhancement of QP DOS due to the minor gap.
Because the magnetic field quickly exceeds the lower critical value of a few millitesla~\cite{Rauchschwalbe1982PRL}, the sample is in the vortex state.
According to calculations based on microscopic quasiclassical theory~\cite{Nakai2002JPSJ,Nakai2004PRB}, 
the initial slope of $C_{\rm e}(H)$ in the vortex state is ${\rm d}C_{\rm e}/{\rm d}H|_{H \sim 0} = [C_{\rm e}(\Hc2)-C_{\rm e}(0)]/(\alpha\Hc2^{\rm orb})$ where $\Hc2^{\rm orb}$ is the orbital-limiting field. 
The parameter $\alpha$ depends on the gap structure, and has a maximum value of  0.8 for an isotropic gap. A minor gap or a gap anisotropy decreases the value of $\alpha$,
eventually approaching zero for a nodal gap.
Using this relation, with $\mu_0\Hc2^{\rm orb}= 10$~T estimated from $\Hc2^{\rm orb} \sim 0.7\Tc{\rm d}\Hc2/{\rm d}T|_{\Tc}$ from Fig.~1(c),
one obtains $\alpha \sim 0.67$ from the $C_{\rm e}(H)$ data at 60~mK for $H \parallel [100]$.
This intermediate value of $\alpha$ favors weakly anisotropic or multiband full-gap superconductivity.

\begin{figure}
\includegraphics[width=3.1in]{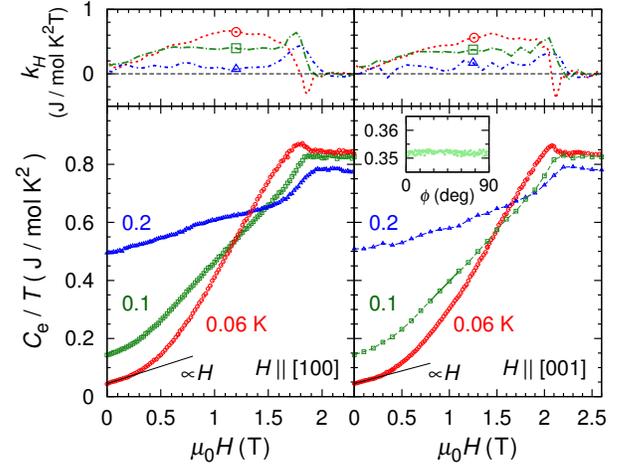}
\caption{
(Color online) 
The ratio $C_{\rm e}/T$ and its slope $k_H={\rm d}(C_{\rm e}/T)/{\rm d}(\mu_0H)$ as a function of magnetic field applied parallel to the [100] and [001] directions at 0.06~(circles), 0.1~(squares), and 0.2~K~(triangles). 
No hysteresis is found with increasing and decreasing fields.
Inset shows $C_{\rm e}/T(\phi)$ at 0.1~K in 0.7~T.
}
\label{sh2}
\end{figure}

To search for the vertical line nodes,
$C_{\rm e}(\phi)$ is measured by rotating the field within the $ab$ plane [see inset of Fig.~2 and Supplemental Material (III)],
where $\phi$ is the azimuthal angle between $H$ and the crystal [100] axis.
Doppler-shift analyses predict that, for a rotating magnetic field, the QP DOS will oscillate and exhibit local minima 
when $H$ is along the nodal or gap-minimum directions~\cite{Vekhter1999PRB,Sakakibara2007JPSJ}.
For example, $C_{\rm e}(\phi)$ for the $d_{x^2-y^2}$-wave superconductor CeIrIn$_5$ (with $\Tc=0.4$~K) has a fourfold oscillation with a large $A_4$ value of 2\%~\cite{Kittaka2012PRB},
where $A_4$ is the amplitude normalized by the field dependent part of the specific heat $C_H=C_{\rm e}(H)-C_{\rm e}(0)$.
In contrast to CeIrIn$_5$, no angular oscillation of $C_{\rm e}(\phi)$ is observed for CeCu$_2$Si$_2$ 
at a temperature of 0.1 (0.2)~K within the 0.1\% (0.5\%) sensitivity of the measurements of $A_4$.
This result implies that QPs are induced by $H$ isotropically with respect to $\phi$, so that vertical line nodes are not detected for $C_{\rm e}(\phi)$.
Possible $C_{\rm e}(\phi)$ oscillation due to the in-plane $\Hc2$ anisotropy \cite{Vieyra2011PRL} was not detected within the present experimental accuracy.

\begin{figure}
\includegraphics[width=3.1in]{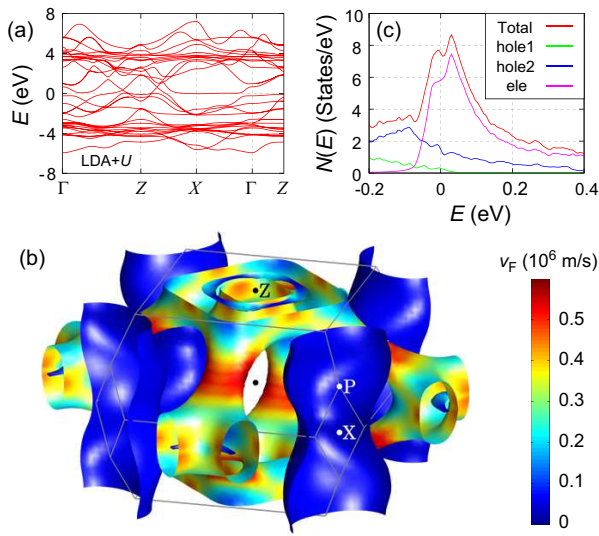} 
\caption{
(Color online) 
(a)~Band structure of CeCu$_2$Si$_2$ calculated using the LDA$+U$ method. 
(b)~The calculated Fermi surfaces colored by the magnitude of the Fermi velocity $v_{\rm F}$. 
(c)~The total density of states and the partial density of states for the three bands.
The Fermi level corresponds to $E=0$.
Here ``hole1'' and ``hole2'' are a small hole ring and a connected hole sheet located around the $Z$ point, respectively, and
``ele'' is a tubular electron sheet located around the $X$ point.
}
\label{FS}
\end{figure}

Based on the present results, 
CeCu$_2$Si$_2$ is a ``nodeless'' multiband superconductor. 
Because $C_{\rm e}(T,H,\phi)$ is sensitive to the contribution from heavy QPs, 
``nodeless'' implies that the gap is fully open in the heavy-mass bands.
To get an insight into the band structure of CeCu$_2$Si$_2$, 
first-principles calculations were performed [see Supplemental Material (IV)]. 
There are a flat electron band around the $X$ point with the heaviest mass and two hole bands around the $Z$ point [Figs.~3(a)-(c)].
The heavy electron band resembles the one obtained in previous studies~\cite{Zwicknagl1993Physica,Eremin2008PRL} and 
its flat parts are connected by the nesting vector $\Vec{Q}=(0.215, 0.215, 1.458)$ around which magnetic excitations have been observed in inelastic neutron experiments~\cite{Stockert2011NatPhy}.

In this Fermi-surface topology, 
the results rule out a $d_{x^2-y^2}$-wave state.
It has line nodes on the electron Fermi sheet of heavy mass,
which is incompatible with ``nodeless'' superconductivity.
Likewise, realization of an ordinary $d_{xy}$-wave state would require quite unusual situations such that 
the effective mass of the hole band is negligibly small so that the nodal structure cannot be detected by the $C(T, H)$ measurements.
Otherwise, we have to seek possibilities of fully-gapped states instead of a nodal $d$-wave state, 
including an unconventional $s$-wave, such as $s_{\pm}$-wave~\cite{Kuroki2008PRL}, 
a conventional $s$-wave, or a fully-gapped $d+{\rm i}d$ state. 
Indeed, the two-gap structure detected in the $C_{\rm e}(T,H)$ measurements indicates that
the mass of the hole bands is not negligible.
Nevertheless, fully-gapped states apparently contradict some key experiments that point to nodal $d$-wave symmetry,
such as the spin resonance observed in neutron scattering~\cite{Stockert2011NatPhy}. 
Therefore, there might remain a possibility of nodal $d$-wave superconductivity,
e.g., with a very unusual evolution of the gap size near the nodes, leading to a small DOS at low energies.
Further investigations are needed to explain this discrepancy. 

%%%%%%% anomalous increase of DOS near Hc2 %%%%%%%

In addition to multiband superconductivity with unexpectedly small QP excitations,
the specific-heat measurements reveal unusual phenomena at high fields.
Above 1.2~T, a remarkable upturn is observed in the $T$ variation of $C_{\rm e}/T$ when cooled below 0.15~K. 
This upturn disappears when $H$ reaches $\Hc2$, indicating that the phenomenon is related to superconductivity. 
Strikingly, $C_{\rm e}/T$ at 1.8~T exceeds the normal-state value when $T \le 80$~mK [Figs.~1(a), 1(c), and 2]
in spite of the presence of a distinct superconducting transition at a higher temperature of about 0.35~K. 
These anomalous features cannot be attributed to the nuclear Schottky contribution that increases in proportion to $H^2$.
Qualitatively the same behavior is observed for the  $H \parallel [001]$ direction as well.

\begin{figure}
\includegraphics[width=3.2in]{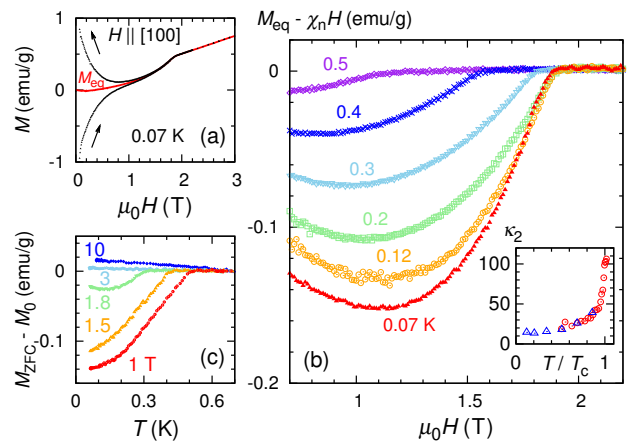}
\caption{
(Color online) 
(a)~Magnetization $M$ of CeCu$_2$Si$_2$ in $H \parallel [100]$ at 70~mK.
The solid line represents the equilibrium magnetization $M_{\rm eq}$ obtained by averaging over increasing and  decreasing field sweeps.
(b)~Graphs of $M_{\rm eq}-\chi_{\rm n}H$ at 0.07, 0.12, 0.2, 0.3, 0.4, and 0.5~K,
where $\chi_{\rm n}H$ is the paramagnetic contribution.
The inset shows temperature dependence of the Maki parameter $\kappa_2$ evaluated from the specific-heat (circles) and the magnetization (triangles)
using the relations ${\rm d}(M_{\rm eq}-\chi_{\rm n}H)/{\rm d}H|_{\Hc2}=1/4\pi\beta(2\kappa_2^2-1)$ and $(\Delta C/T)|_{\Tc}=({\rm d}\Hc2/{\rm d}T)^2/4\pi\beta(2\kappa_2^2-1)$.
Here, $\Delta C$ is the jump in the specific heat at $\Tc(H)$ from Fig.~1(a). The value $\beta=1.16$ assumes a triangular vortex lattice.
The slope of ${\rm d}\Hc2/{\rm d}T$ is estimated using the data in Fig.~1(c).
(c)~Magnetization measured during a zero-field cooling process, $M_{\rm ZFC}$, at 1, 1.5, 1.8, 3, and 10~T
after subtracting the value $M_0$ of $M_{\rm ZFC}$ at 0.7~K.
}
\label{M}
\end{figure}

To further investigate this strange high-field behavior, 
the dc magnetization $M(T,H)$ is measured down to 70~mK for $H \parallel [100]$ using the same sample.
At 70~mK, the hysteresis loop of $M(H)$ is small except at low fields [Fig.~4(a)]. 
This result verifies the purity of the sample.
A diamagnetic contribution can be observed up to 1.9~T, implying that the sample remains superconducting even when $C_{\rm e}/T$ exceeds its normal-state value.
Nevertheless, the diamagnetic contribution in the range  1.8~T $\lesssim \mu_0H \le 1.9$~T is unusually suppressed upon cooling below 0.12~K [Figs.~4(b) and 4(c)],
suggesting a strong pair-breaking effect near $\Hc2$.
Indeed, a kink develops in $M(H)$ at $\Hc2$ when cooled, attributed to a strong Pauli paramagnetic effect~\cite{Ichioka2007PRB,Machida2008PRB}. 
This effect is also evident in $C_{\rm e}(H)$ for $T=$~0.1 and 0.2~K.
The slope of $C_{\rm e}(H)$, i.e., $k_H={\rm d}(C_{\rm e}/T)/{\rm d}(\mu_0H)$, is enhanced as $H$ approaches $\Hc2$ (Fig.~2), 
although the behavior is hidden at 60~mK by the anomalous $C_{\rm e}/T$ upturn in the high-field state.
The Maki parameter $\kappa_2$ shows a large decrease on cooling near $\Tc$ [inset of Fig.~4(b)], 
in good agreement with theory for strongly Pauli-limited superconductors.
It is concluded that the unusual high-field behavior can be ascribed to a strong paramagnetic depairing effect.

For a clean superconductor with a strong paramagnetic effect, 
the superconducting-to-normal phase transition is expected to change from second to first order at low temperatures~\cite{Matsuda2007JPSJ}. 
However, the transition at $\Hc2$ oddly remains second order in CeCu$_2$Si$_2$, indicated by 
a continuous change of $M(H)$ across $\Hc2$ without hysteresis and a leveling off of $\kappa_2$ at low temperatures.
These features cannot be explained simply by the paramagnetic effect for single-band superconductivity.

Returning to the $T$ variation of $C_{\rm e}/T$ in Fig.~1(a), 
it can be shown that its unusual high-field behavior is related to the energy dependence of the total DOS.
In general,
\begin{equation}
C_{\rm e}/T=\int x^2N(xT)/[4\cosh^2(x/2)]{\rm d}x,
\end{equation}
where $N(E)$ is the energy dependence of the total DOS and 
the energy $E$ is replaced with $xT$.
By approximating $N(E)=a|E|^n$, which is valid near $E \sim 0$,
Eq.~(1) can be rewritten as $C_{\rm e}/T=a^\prime N(T)$,
where $a$, $n$, and $a^\prime$ are constants.
This relationship demonstrates that the $T$ dependence of $C_{\rm e}/T$ at low temperatures mimics $N(E)$ at low energies.

In this context, the $T$ variation in $C_{\rm e}/T$ at each field can be understood 
by taking into account the presence of two gaps as well as the strong paramagnetic effect.
Theory predicts that
$N(E)$ has a V-shaped structure in the vortex state~\cite{Nakai2006PRB}, 
i.e., $N(E) \propto |E|$ near $E \sim 0$ with an edge-singularity peak at $|E| \sim \Delta$.
The observed kink in $C_{\rm e}/T$ at low $H$, such as at 0.6~T, is ascribed to the edge singularity of the small V-shaped DOS of a minor gap 
superposed on the large V-shaped DOS of a major gap. 
At higher fields, assuming the strong paramagnetic effect holds for the minor gap, 
the edge singularity is shifted toward lower energy~\cite{Ichioka2007PRB} and 
an upturn in $C_{\rm e}/T$ corresponding to the tail of the singularity in the high-energy side is observed.
Because the V-shaped DOS of the major gap gradually approaches the normal-state DOS with increasing field, 
$C_{\rm e}/T$ can exceed the normal-state value near $\Hc2$ 
if the DOS enhancement due to the edge singularity of the minor gap remains prominent.
The upward curvature in $C_{\rm e}(H)$ near 0.5~T is a sign of the paramagnetic effect for the minor gap.
While these analyses suggest that the two-band full-gap model, in the presence of a strong paramagnetic effect, can explain the anomalous $C_{\rm e}(T,H)$ behavior satisfactorily, 
there remain other possible origins, such as the occurrence of an AFM ordering in the high-field superconducting state. 
To further confirm the Pauli-limited two-gap scenario, detailed calculations of $C(T,H)$ based on the microscopic theory are in progress \cite{Machida}.

In summary, we have investigated the low-temperature specific heat and magnetization of a high-quality \textit{S}-type single crystal of CeCu$_2$Si$_2$. 
Our study has provided thermodynamic evidence for multiband superconductivity, 
an unexpected deficiency of nodal QP excitations, and a strong Pauli paramagnetic effect in CeCu$_2$Si$_2$.
The discovery of unexpectedly small QP DOS at low energies 
challenges the long-held view of this heavy-fermion superconductor whose pairing symmetry is believed to be of the nodal $d$-wave type. 
These findings help resolve long-standing issues about the pairing mechanism in CeCu$_2$Si$_2$.

We acknowledge helpful discussions with Y. Kitaoka.
We also thank Y. Tsutsumi for supporting the calculation on the basis of the two-gap model.
H. I. thanks M. -T. Suzuki for assistance with the LDA$+U$ calculations.
This work was supported by a Grants-in-Aid for Scientific Research on Innovative Areas ``Heavy Electrons'' (20102007, 23102705)
from MEXT, and KAKENHI (25800186, 24340075, 21340103, 23340095, 24540369) from JSPS.

\clearpage
\onecolumngrid
\appendix

\begin{center}
{\large Supplemental Material for \\
\textbf{Multiband Superconductivity with Unexpected Deficiency of Nodal Quasiparticles in CeCu$_2$Si$_2$}}\\
\vspace{0.1in}
Shunichiro Kittaka,$^{1}$ Yuya Aoki,$^{1}$ Yasuyuki Shimura,$^{1}$ Toshiro Sakakibara,$^{1}$\\ Silvia Seiro,$^2$ Christoph Geibel,$^2$ Frank Steglich,$^2$ Hiroaki Ikeda,$^3$ and Kazushige Machida$^4$\\
{\small 
\textit{$^1$Institute for Solid State Physics, University of Tokyo, Kashiwa, Chiba 277-8581, Japan}\\
\textit{$^2$Max Planck Institute for Chemical Physics of Solids, 01187 Dresden, Germany}\\
\textit{$^3$Department of Physics, Kyoto University, Kyoto 606-8502, Japan}\\
\textit{$^4$Department of Physics, Okayama University, Okayama 700-8530, Japan}\\
}
(Dated: \today)
\end{center}

\section{I. Nuclear specific heat}
The total specific heat includes electronic, phonon, and nuclear contributions.
To explore the variation of the quasiparticle density of states, which is proportional to the electronic part of the specific heat at low temperatures, 
subtraction of the other contributions is necessary. 
Figure~\ref{Supp}(a) plots the total specific heat of an \textit{S}-type CeCu$_2$Si$_2$ single crystal divided by temperature, $C/T$,
measured with various magnetic fields applied along the [100] axis. 
In this temperature range, the phonon contribution is negligible.
By contrast, the contribution from the $^{63, 65}$Cu ($I=3/2$) nuclei cannot be neglected at high fields:
a striking upturn in $C/T$ appears upon cooling.

To evaluate the nuclear Schottky contribution, 
the field variation, $a(H)$, in the coefficient of the $1/T^2$ term in $C(T)$ is found by fitting the normal-state data at 1.9, 3, 4, and 5~T between 50~mK~$\le T \le 75$~mK 
using the function $C_{\rm n}=a(H)/T^2+\gamma T$ 
[see the lower inset of Fig.~\ref{Supp}(a)]. 
The result is graphed as the circles in the upper inset of Fig.~\ref{Supp}(a) and is $a(H) \simeq 7.4H^2$~$\mu$J$\cdot$K/mol (dashed line),
which is comparable to the calculation, $a(H)=(6.4H^2+0.1)$~$\mu$J$\cdot$K/mol (solid line), using a nuclear spin Hamiltonian. 
In this study we define $C_{\rm n}=(7.4H^2+0.1)/T^2$~$\mu$J$\cdot$/(mol K) so that $C_{\rm e}/T$ at 60~mK becomes constant above $\Hc2$
[see the inset of Fig.~\ref{Supp}(b)]. 
The $C/T$ data after subtracting this $C_{\rm n}$ contribution is graphed in Fig.~\ref{Supp}(b).
The nuclear contribution is negligible at zero field.
Note that the key features in this study are evident in Fig.~\ref{Supp}(a):
(i)~an exponential $T$ dependence of $C(T)$ at 0~T in good agreement with a two-gap model,
(ii)~the absence of a $\sqrt{H}$ variation in $C(H)$, 
(iii)~a kink in the temperature variation of $C/T$ at 0.6~T, which suggests the presence of a minor gap, and
(iv)~a low-temperature enhancement of $C/T$ at 1.8~T.
These facts ensure that the error in the subtraction of the nuclear specific heat does not affect the conclusion in our study. 

\section{II. Analyses of the low-temperature specific-heat data}
In general, 
$C_{\rm e}(T)$ exhibits exponential and power-law $T$ dependence at low temperatures for fully-gapped and nodal superconductors, respectively, 
reflecting the low-energy quasiparticle excitations across the gap.
As shown in Fig.~\ref{lowTfit}, we fit the low-temperature part of the zero-field $C_{\rm e}(T)$ data in the range 0.04~K~$\le T \le 0.15$~K (dashed lines)
using the power-law functions, (a) $C=\beta_0T^3+\gamma_0T$ and (b) $C=\beta_0T^n+\gamma_0T$, and the BCS function, (c) $C=A\exp(-\Delta_0/T)+\gamma_0T$, 
with adjustable parameters $\beta_0$, $\gamma_0$, $n$, $A$, and $\Delta_0$.
It is obvious that the power-law functions do not match the experimental data, whereas the BCS function gives the best fit down to 40~mK.

However, it is possible that the impurity-scattering effect enhances the residual DOS and smears out the power-law behavior close to 0~K.
To examine this possibility, we fit the data in the range 0.06~K~$\le T \le 0.15$~K, as well (dotted lines in Fig.~\ref{lowTfit}).
Although the accuracy of the power-law fits is slightly improved in that temperature range, 
the residual specific-heat value at 0~K, $\gamma_0$, unfavorably becomes negative.
This again supports that the power-law fit to $C_{\rm e}(T)$ is not valid for CeCu$_2$Si$_2$.
These analyses along with the two-gap fit [Fig.~1(b)] suggest that node is absent on both the heaviest and second-heaviest bands.

If one supposes that CeCu$_2$Si$_2$ has the Fermi-surface topology described in Fig.~3, 
major and minor gaps detected from $C_{\rm e}(T)$ correspond to the gaps opening on the electron and ``hole2'' bands, respectively.
In this case, both the $d_{x^2-y^2}$ and $d_{xy}$ symmetries are unfavorable because they have line nodes on the ``hole2'' band.
Indeed, as demonstrated in Fig.~\ref{twogap}, 
the best fit using a major full gap and a minor line-node gap (for simplicity, a spherical Fermi surface and a $d_{xy}$-wave gap are assumed), 
whose gap sizes are $1.3k_{\rm B}\Tc$ and $2.7k_{\rm B}\Tc$ and weights are 70\% and 30\% of the total DOS, respectively, 
gives worse results compared with the two-full-gap fit, particularly in the low-$T$ region.

\section{III. Field-angle-resolved specific heat}
Figure \ref{Cphi} shows $C_{\rm e}/T(\phi)$ of CeCu$_2$Si$_2$ measured in a rotating magnetic field within the $ab$ plane. 
The angle $\phi$ denotes the azimuthal field angle measured from the [100] axis.
The dashed lines in Fig.~\ref{Cphi} are fits to the data using $C_{\rm e}(\phi)=C_0+C_H[1+A_4\cos(4\phi)]$,
where $C_0$ and $C_H$ are the zero-field and field-dependent parts of $C_{\rm e}$, respectively, and 
$A_4$ is the amplitude of the fourfold oscillation normalized by $C_H$.
Although $C_{\rm e}(\phi)$ was investigated at various $H$ and $T$,
no remarkable oscillation was detected in $C_{\rm e}(\phi)$ within an experimental error.

\section{IV. Electronic band structure}
Relativistic electronic structure calculations were performed using the full-potential augmented plane-wave plus local orbital method 
including spin-orbit coupling (APW-LO), as implemented in the WIEN2K code \cite{rf:Blaha}. 
For the exchange-correlation functional, the Perdew-Burke-Ernzerhof form \cite{rf:Perdew} was adopted. 
The crystal structure is body-centered tetragonal (space group: No.~139, I4/mmm).  
The lattice parameters and atomic positions are taken from experimental data \cite{rf:Jarlborg}. 
The muffin-tin radii $R_{\rm MT}$ of Ce, Cu, and Si were taken to be 2.50, 2.41, and 2.13 Bohr, respectively, 
and the maximum modulus of the reciprocal vectors $K_{\rm max}$ was chosen such that $R_{\rm MT} K_{\rm max} = 7.0$. 
The first Brillouin zone was sampled using a $11\times 11\times 11$ $k$-mesh.

Figure~\ref{LDAband}(a) depicts the electronic band structure along the high-symmetry line. 
The compound is a compensated metal with $f$-electron number $n_f\simeq 1.05$. Three bands cross the Fermi level at $E=0$. 
Figure~\ref{LDAband}(b) plots the density of states near the Fermi level. In Figs.~\ref{LDAband}(c)-(e), 
the Fermi surface is colored by the Fermi velocity. 
There appear a connected hole sheet, a small pocket around the $Z$ point, a cubic electron sheet around the $\Gamma$ point, and a tiny cigar pocket. 
These results are consistent with the previous study \cite{rf:Harima}.

Next, a tight-binding Hamiltonian is constructed by downfolding the APW-LO Hamiltonian using the wannier90 code \cite{rf:Marzari,rf:Souza,rf:Mostofi} 
via the wien2wannier interface \cite{rf:Kunes}, based on 56 Wannier orbitals (Ce $5d$ and $4f$, Cu $3d$, and Si $3p$ orbitals). 
In light of the on-site Coulomb interactions between $4f$ electrons, the Hatree-Fock approximation is applied with fixed $n_f$, 
corresponding to the conventional LDA$+U$ method \cite{rf:Anisimov,rf:Suzuki}. 
The empirical values $U=U'=7$~eV and $J=0$~eV were adopted \cite{rf:Kang}.

The obtained $+U$ band structure is illustrated in Fig.~3(a) of the main text. 
As expected, some empty $f$-electron bands are shifted up by $\sim 3.5$~eV. 
A flat band appears near the $X$ point. 
Figures~\ref{LDAUband}(a)-(c) are the corresponding Fermi surfaces, composed of 
an electron sheet around the $X$ point with a heavy mass, a connected hole sheet with a light mass, and a small hole ring. 
The overall structure is similar to the renormalized band of Zwicknagl \cite{rf:Zwicknagl}. 
As discussed in Ref.~\onlinecite{rf:Eremin}, 
a nesting property in the tubular electron sheet is compatible with the incommensurate {\bf Q} vector observed in neutron scattering measurements.

\clearpage

\setcounter{figure}{0}
\renewcommand{\thefigure}{S\arabic{figure}}
\begin{figure}
\includegraphics[width=5.5in]{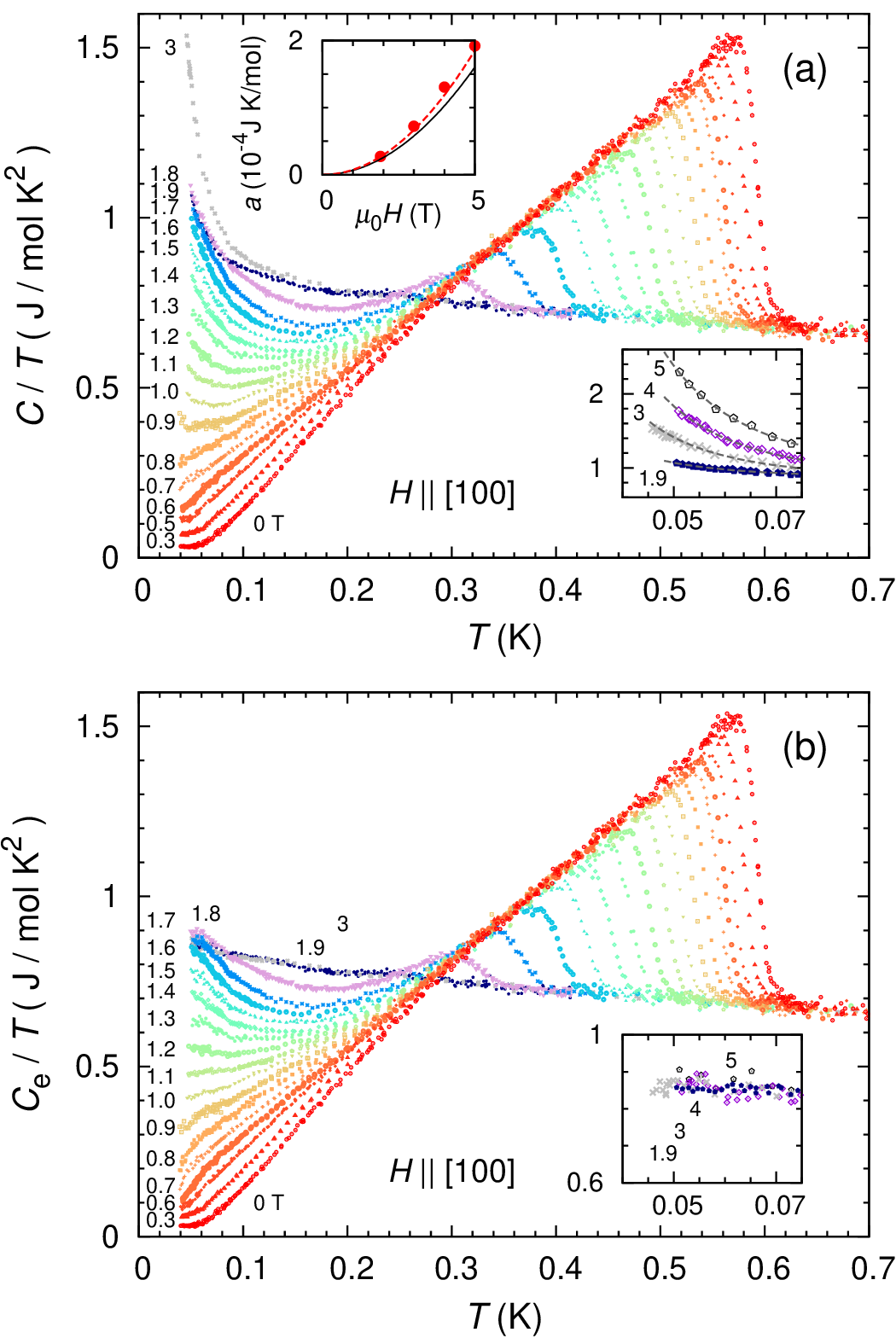} 
\caption{
(a) Total specific heat of an \textit{S}-type CeCu$_2$Si$_2$ divided by $T$, $C/T$, as a function of temperature measured 
in magnetic fields of 0, 0.3, 0.5, 0.6, 0.7, 0.8, 0.9, 1, 1.1, 1.2, 1.3, 1.4, 1.5, 1.6, 1.7, 1.8, 1.9, and 3~T (from bottom to top at $T=0.15$~K) applied along the [100] direction.
(b) The same data after subtracting the nuclear specific heat contribution $C_{\rm n}$. 
The lower inset in (a) shows $C/T$ at high fields of 1.9, 3, 4, and 5~T and the fits using the function $C_{\rm n}=a(H)/T^2+\gamma T$ (dashed lines).
The upper inset in (a) plots the field variation in the coefficient of the $T^{-2}$ term in the $C(T)$ data, $a(H)$, as the circles. 
The dashed line is the function $a(H)$ used in the present study and 
the solid line is a calculation using a nuclear spin Hamiltonian.
The inset in (b) shows $C_{\rm e}/T$ at high fields of 1.9, 3, 4, and 5~T.
}
\label{Supp}
\end{figure}

\begin{figure}
\includegraphics[width=5.in]{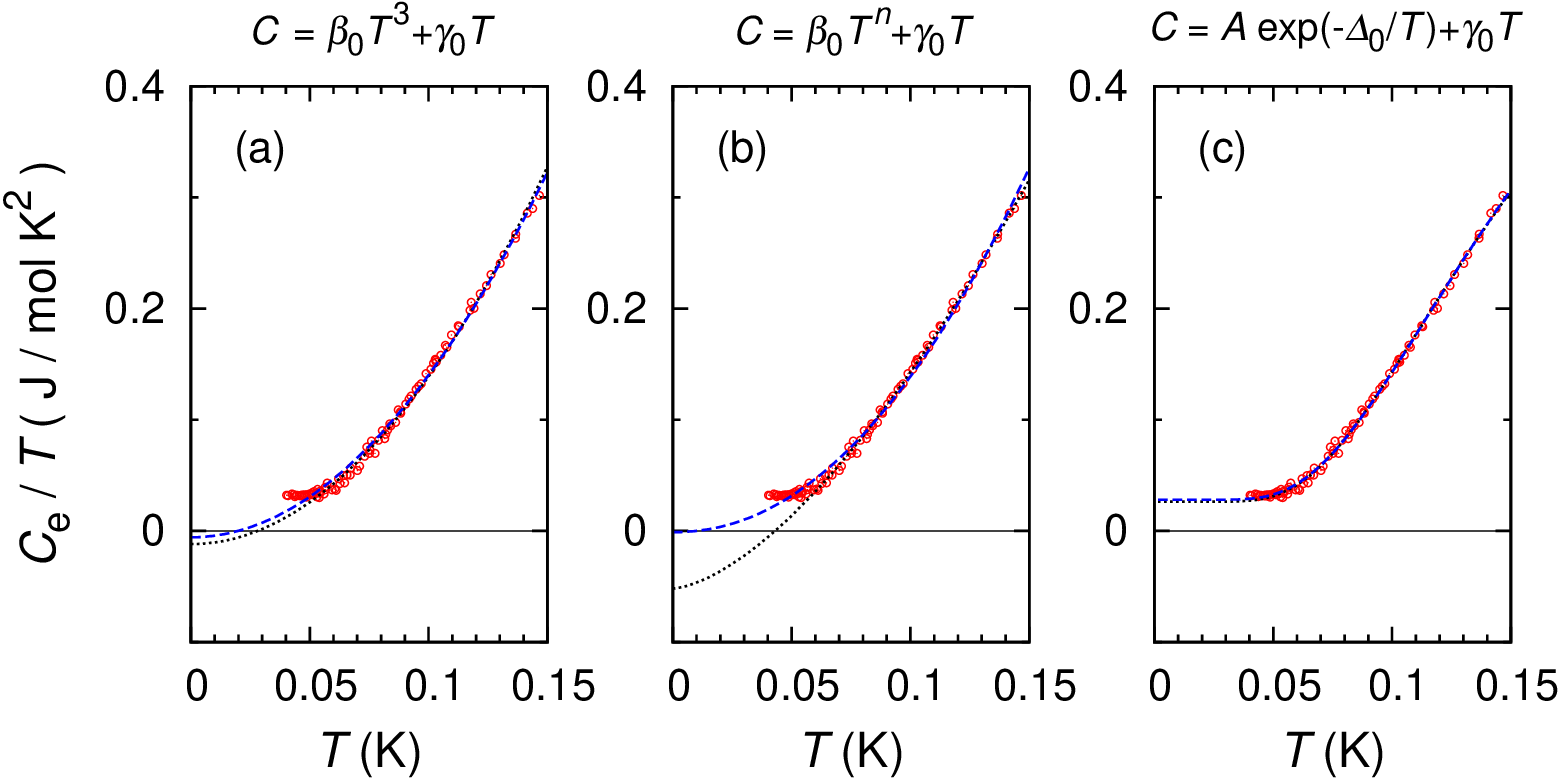} 
\caption{
Low-temperature part of the specific-heat data at zero field compared with the fitting results using functions (a) $C=\beta_0T^3+\gamma_0T$, (b) $C=\beta_0T^n+\gamma_0T$, and (c) $C=A\exp(-\Delta_0/T)+\gamma_0T$.
The dashed and dotted lines are the results obtained by fitting the data in the range 0.04~K~$\le T \le 0.15$~K and 0.06~K~$\le T \le 0.15$~K, respectively. 
}
\label{lowTfit}
\end{figure}

\begin{figure}
\includegraphics[width=4.2in]{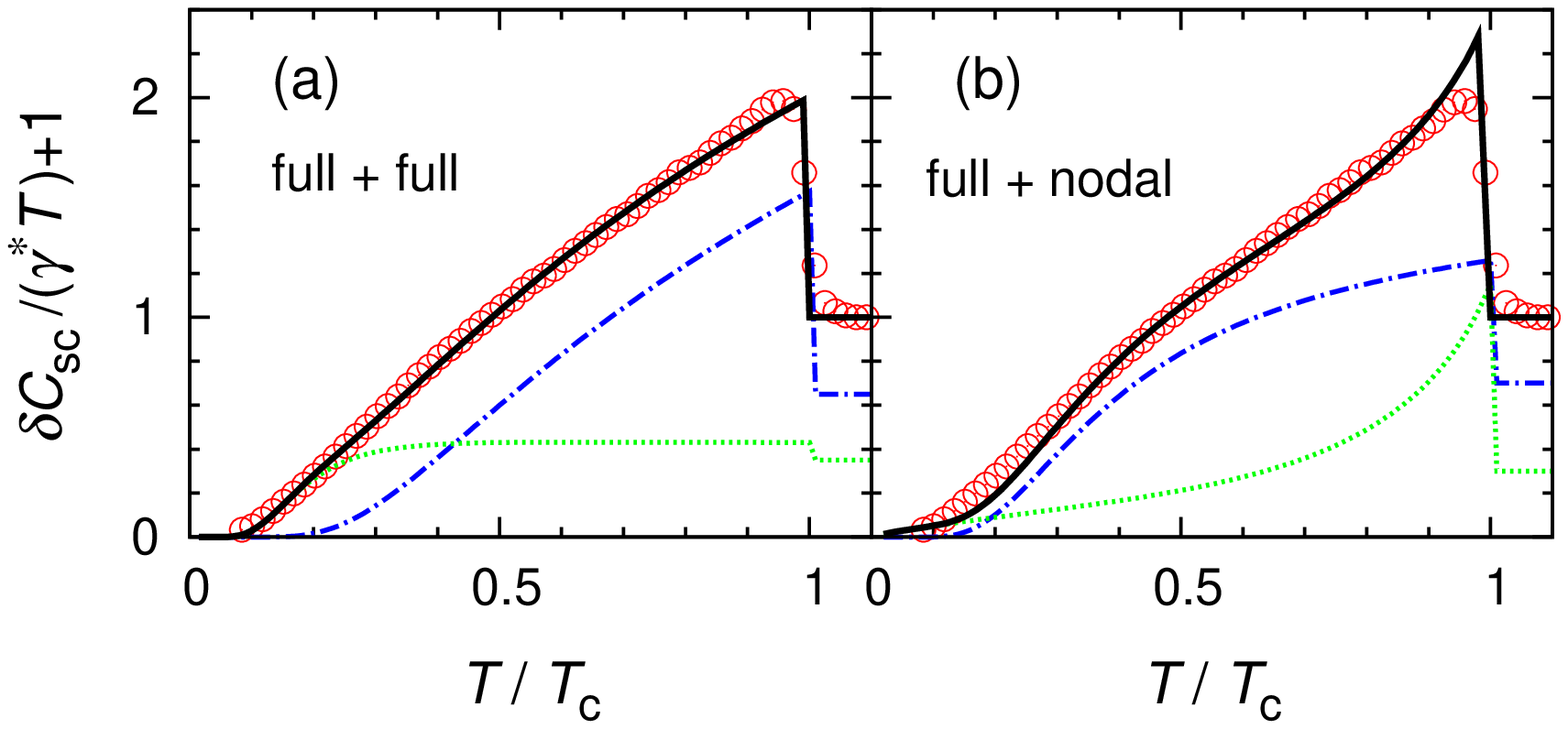} 
\caption{
Temperature variation of $\delta C_{\rm sc} (T, H)/\gamma^\ast T+1$ at $H=0$ (circles) compared with the best fit on the basis of the two-gap model (solid line), 
using (a) two full gaps and (b) major full and minor line-node gaps. Contribution of each gap to the total specific heat is represented by the broken line.
}
\label{twogap}
\end{figure}

\begin{figure}
\includegraphics[width=3.4in]{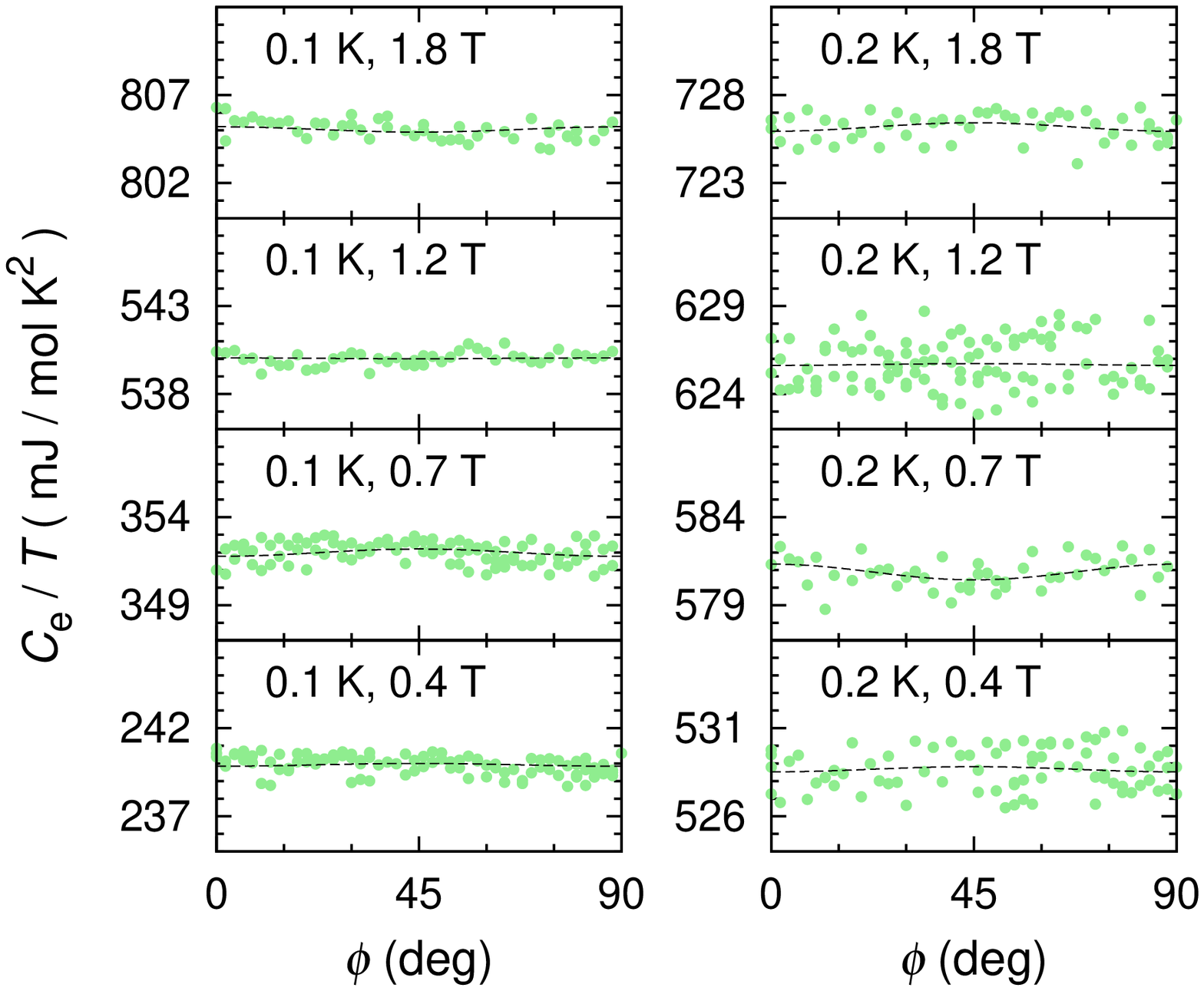}
\caption{
Field-angle-resolved $C_{\rm e}/T$ at 0.1 and 0.2~K in a magnetic field rotated within the $ab$ plane 
as a function of the field angle $\phi$ measured relative to the [100] axis.
}
\label{Cphi}
\end{figure}

\begin{figure}
\includegraphics[width=\textwidth]{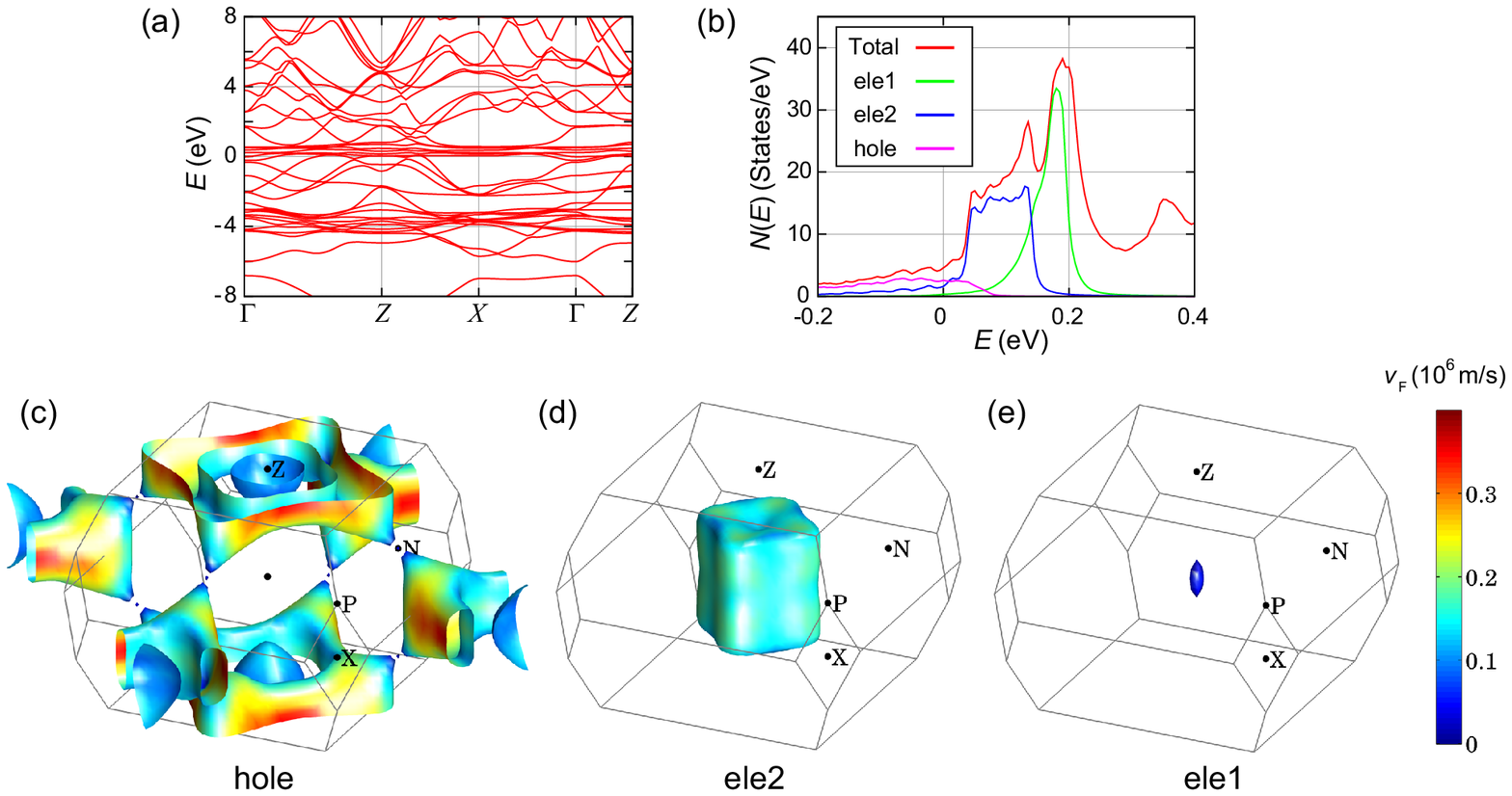}
\caption{
(a) Band structure along the high-symmetry line. The Fermi level is located at $E=0$.  
(b) The partial density of states for three bands across the Fermi level. 
(c)-(e) The corresponding Fermi surfaces colored by the Fermi velocity.}
\label{LDAband}
\end{figure}
\begin{figure}
\includegraphics[width=\textwidth]{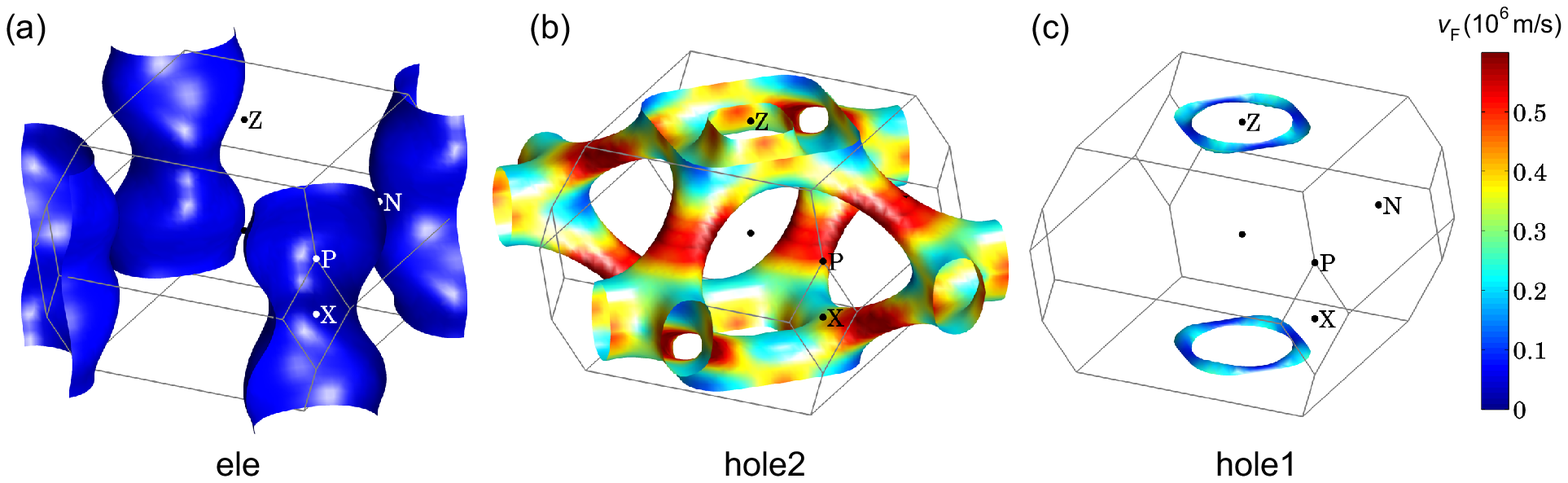}
\caption{
(a)-(c) Fermi surfaces calculated by the LDA$+U$ method. 
The appearance of an electron sheet with a heavy mass [panel (a)] is consistent with the trend in the renormalized band of Zwicknagl. }
\label{LDAUband}
\end{figure}
%\bibliography{C:/usr/local/share/texmf/bibref/ref_CeCu2Si2.bib}

\begin{thebibliography}{28}
\expandafter\ifx\csname natexlab\endcsname\relax\def\natexlab#1{#1}\fi
\expandafter\ifx\csname bibnamefont\endcsname\relax
  \def\bibnamefont#1{#1}\fi
\expandafter\ifx\csname bibfnamefont\endcsname\relax
  \def\bibfnamefont#1{#1}\fi
\expandafter\ifx\csname citenamefont\endcsname\relax
  \def\citenamefont#1{#1}\fi
\expandafter\ifx\csname url\endcsname\relax
  \def\url#1{\texttt{#1}}\fi
\expandafter\ifx\csname urlprefix\endcsname\relax\def\urlprefix{URL }\fi
\providecommand{\bibinfo}[2]{#2}
\providecommand{\eprint}[2][]{\url{#2}}
%
\bibitem[{\citenamefont{Steglich et~al.}(1979)\citenamefont{Steglich, Aarts,
  Bredl, Lieke, Meschede, Franz, and Sch\"afer}}]{Steglich1979PRL}
\bibinfo{author}{\bibfnamefont{F.}~\bibnamefont{Steglich}},
  \bibinfo{author}{\bibfnamefont{J.}~\bibnamefont{Aarts}},
  \bibinfo{author}{\bibfnamefont{C.~D.} \bibnamefont{Bredl}},
  \bibinfo{author}{\bibfnamefont{W.}~\bibnamefont{Lieke}},
  \bibinfo{author}{\bibfnamefont{D.}~\bibnamefont{Meschede}},
  \bibinfo{author}{\bibfnamefont{W.}~\bibnamefont{Franz}}, \bibnamefont{and}
  \bibinfo{author}{\bibfnamefont{H.}~\bibnamefont{Sch\"afer}},
  \bibinfo{journal}{Phys. Rev. Lett.} \textbf{\bibinfo{volume}{43}},
  \bibinfo{pages}{1892} (\bibinfo{year}{1979}).
%
\bibitem[{\citenamefont{Stockert et~al.}(2011)\citenamefont{Stockert, Arndt,
  Faulhaber, Geibel, Jeevan, Kirchner, Loewenhaupt, Schmalzl, Schmidt, Si
  et~al.}}]{Stockert2011NatPhy}
\bibinfo{author}{\bibfnamefont{O.}~\bibnamefont{Stockert}},
  \bibinfo{author}{\bibfnamefont{J.}~\bibnamefont{Arndt}},
  \bibinfo{author}{\bibfnamefont{E.}~\bibnamefont{Faulhaber}},
  \bibinfo{author}{\bibfnamefont{C.}~\bibnamefont{Geibel}},
  \bibinfo{author}{\bibfnamefont{H.~S.} \bibnamefont{Jeevan}},
  \bibinfo{author}{\bibfnamefont{S.}~\bibnamefont{Kirchner}},
  \bibinfo{author}{\bibfnamefont{M.}~\bibnamefont{Loewenhaupt}},
  \bibinfo{author}{\bibfnamefont{K.}~\bibnamefont{Schmalzl}},
  \bibinfo{author}{\bibfnamefont{W.}~\bibnamefont{Schmidt}},
  \bibinfo{author}{\bibfnamefont{Q.}~\bibnamefont{Si}}, \bibnamefont{and}
  \bibinfo{author}{\bibfnamefont{F.}~\bibnamefont{Steglich}},
  \bibinfo{journal}{Nature Phys.} \textbf{\bibinfo{volume}{7}},
  \bibinfo{pages}{119} (\bibinfo{year}{2011}).
%
\bibitem[{\citenamefont{Ueda et~al.}(1987)\citenamefont{Ueda, Kitaoka, Yamada,
  Kohori, Kohara, and Asayama}}]{Ueda1987JPSJ}
\bibinfo{author}{\bibfnamefont{K.}~\bibnamefont{Ueda}},
  \bibinfo{author}{\bibfnamefont{Y.}~\bibnamefont{Kitaoka}},
  \bibinfo{author}{\bibfnamefont{H.}~\bibnamefont{Yamada}},
  \bibinfo{author}{\bibfnamefont{Y.}~\bibnamefont{Kohori}},
  \bibinfo{author}{\bibfnamefont{T.}~\bibnamefont{Kohara}}, \bibnamefont{and}
  \bibinfo{author}{\bibfnamefont{K.}~\bibnamefont{Asayama}},
  \bibinfo{journal}{J. Phys. Soc. Jpn.} \textbf{\bibinfo{volume}{56}},
  \bibinfo{pages}{867} (\bibinfo{year}{1987}).
%
\bibitem[{\citenamefont{Kitaoka et~al.}(1986)\citenamefont{Kitaoka, Ueda,
  Fujiwara, Arimoto, Iida, and Asayama}}]{Kitaoka1986JPSJ}
\bibinfo{author}{\bibfnamefont{Y.}~\bibnamefont{Kitaoka}},
  \bibinfo{author}{\bibfnamefont{K.}~\bibnamefont{Ueda}},
  \bibinfo{author}{\bibfnamefont{K.}~\bibnamefont{Fujiwara}},
  \bibinfo{author}{\bibfnamefont{H.}~\bibnamefont{Arimoto}},
  \bibinfo{author}{\bibfnamefont{H.}~\bibnamefont{Iida}}, \bibnamefont{and}
  \bibinfo{author}{\bibfnamefont{K.}~\bibnamefont{Asayama}},
  \bibinfo{journal}{J. Phys. Soc. Jpn.} \textbf{\bibinfo{volume}{55}},
  \bibinfo{pages}{723} (\bibinfo{year}{1986}).
%
\bibitem[{\citenamefont{Ishida et~al.}(1999)\citenamefont{Ishida, Kawasaki,
  Tabuchi, Kashima, Kitaoka, Asayama, Geibel, and Steglich}}]{Ishida1999PRL}
\bibinfo{author}{\bibfnamefont{K.}~\bibnamefont{Ishida}},
  \bibinfo{author}{\bibfnamefont{Y.}~\bibnamefont{Kawasaki}},
  \bibinfo{author}{\bibfnamefont{K.}~\bibnamefont{Tabuchi}},
  \bibinfo{author}{\bibfnamefont{K.}~\bibnamefont{Kashima}},
  \bibinfo{author}{\bibfnamefont{Y.}~\bibnamefont{Kitaoka}},
  \bibinfo{author}{\bibfnamefont{K.}~\bibnamefont{Asayama}},
  \bibinfo{author}{\bibfnamefont{C.}~\bibnamefont{Geibel}}, \bibnamefont{and}
  \bibinfo{author}{\bibfnamefont{F.}~\bibnamefont{Steglich}},
  \bibinfo{journal}{Phys. Rev. Lett.} \textbf{\bibinfo{volume}{82}},
  \bibinfo{pages}{5353} (\bibinfo{year}{1999}).
%
\bibitem[{\citenamefont{Fujiwara et~al.}(2008)\citenamefont{Fujiwara, Hata,
  Kobayashi, Miyoshi, Takeuchi, Shimaoka, Kotegawa, Kobayashi, Geibel, and
  Steglich}}]{Fujiwara2008JPSJ}
\bibinfo{author}{\bibfnamefont{K.}~\bibnamefont{Fujiwara}},
  \bibinfo{author}{\bibfnamefont{Y.}~\bibnamefont{Hata}},
  \bibinfo{author}{\bibfnamefont{K.}~\bibnamefont{Kobayashi}},
  \bibinfo{author}{\bibfnamefont{K.}~\bibnamefont{Miyoshi}},
  \bibinfo{author}{\bibfnamefont{J.}~\bibnamefont{Takeuchi}},
  \bibinfo{author}{\bibfnamefont{Y.}~\bibnamefont{Shimaoka}},
  \bibinfo{author}{\bibfnamefont{H.}~\bibnamefont{Kotegawa}},
  \bibinfo{author}{\bibfnamefont{T.~C.} \bibnamefont{Kobayashi}},
  \bibinfo{author}{\bibfnamefont{C.}~\bibnamefont{Geibel}}, \bibnamefont{and}
  \bibinfo{author}{\bibfnamefont{F.}~\bibnamefont{Steglich}},
  \bibinfo{journal}{J. Phys. Soc. Jpn.} \textbf{\bibinfo{volume}{77}},
  \bibinfo{pages}{123711} (\bibinfo{year}{2008}).
%
\bibitem[{\citenamefont{Vieyra et~al.}(2011)\citenamefont{Vieyra, Oeschler,
  Seiro, Jeevan, Geibel, Parker, and Steglich}}]{Vieyra2011PRL}
\bibinfo{author}{\bibfnamefont{H.~A.} \bibnamefont{Vieyra}},
  \bibinfo{author}{\bibfnamefont{N.}~\bibnamefont{Oeschler}},
  \bibinfo{author}{\bibfnamefont{S.}~\bibnamefont{Seiro}},
  \bibinfo{author}{\bibfnamefont{H.~S.} \bibnamefont{Jeevan}},
  \bibinfo{author}{\bibfnamefont{C.}~\bibnamefont{Geibel}},
  \bibinfo{author}{\bibfnamefont{D.}~\bibnamefont{Parker}}, \bibnamefont{and}
  \bibinfo{author}{\bibfnamefont{F.}~\bibnamefont{Steglich}},
  \bibinfo{journal}{Phys. Rev. Lett.\label{Vieyra2011PRL}}
  \textbf{\bibinfo{volume}{106}}, \bibinfo{pages}{207001}
  (\bibinfo{year}{2011}).
%
\bibitem[{\citenamefont{Eremin et~al.}(2008)\citenamefont{Eremin, Zwicknagl,
  Thalmeier, and Fulde}}]{Eremin2008PRL}
\bibinfo{author}{\bibfnamefont{I.}~\bibnamefont{Eremin}},
  \bibinfo{author}{\bibfnamefont{G.}~\bibnamefont{Zwicknagl}},
  \bibinfo{author}{\bibfnamefont{P.}~\bibnamefont{Thalmeier}},
  \bibnamefont{and} \bibinfo{author}{\bibfnamefont{P.}~\bibnamefont{Fulde}},
  \bibinfo{journal}{Phys. Rev. Lett.} \textbf{\bibinfo{volume}{101}},
  \bibinfo{pages}{187001} (\bibinfo{year}{2008}).
%
\bibitem[{\citenamefont{Seiro et~al.}(2010)\citenamefont{Seiro, Deppe, Jeevan,
  Burkhardt, and Geibel}}]{Seiro2010PSSB}
\bibinfo{author}{\bibfnamefont{S.}~\bibnamefont{Seiro}},
  \bibinfo{author}{\bibfnamefont{M.}~\bibnamefont{Deppe}},
  \bibinfo{author}{\bibfnamefont{H.}~\bibnamefont{Jeevan}},
  \bibinfo{author}{\bibfnamefont{U.}~\bibnamefont{Burkhardt}},
  \bibnamefont{and} \bibinfo{author}{\bibfnamefont{C.}~\bibnamefont{Geibel}},
  \bibinfo{journal}{Phys. Status Solidi B \label{Seiro2010PSSB}}
  \textbf{\bibinfo{volume}{247}}, \bibinfo{pages}{614} (\bibinfo{year}{2010}).
%
\bibitem[{\citenamefont{Gegenwart et~al.}(1998)\citenamefont{Gegenwart,
  Langhammer, Geibel, Helfrich, Lang, Sparn, Steglich, Horn, Donnevert, Link
  et~al.}}]{Gegenwart1998}
\bibinfo{author}{\bibfnamefont{P.}~\bibnamefont{Gegenwart}},
  \bibinfo{author}{\bibfnamefont{C.}~\bibnamefont{Langhammer}},
  \bibinfo{author}{\bibfnamefont{C.}~\bibnamefont{Geibel}},
  \bibinfo{author}{\bibfnamefont{R.}~\bibnamefont{Helfrich}},
  \bibinfo{author}{\bibfnamefont{M.}~\bibnamefont{Lang}},
  \bibinfo{author}{\bibfnamefont{G.}~\bibnamefont{Sparn}},
  \bibinfo{author}{\bibfnamefont{F.}~\bibnamefont{Steglich}},
  \bibinfo{author}{\bibfnamefont{R.}~\bibnamefont{Horn}},
  \bibinfo{author}{\bibfnamefont{L.}~\bibnamefont{Donnevert}},
  \bibinfo{author}{\bibfnamefont{A.}~\bibnamefont{Link}}, \bibnamefont{and}
  \bibinfo{author}{\bibfnamefont{W.}~\bibnamefont{Assmus}},
  \bibinfo{journal}{Phys. Rev. Lett.} \textbf{\bibinfo{volume}{81}},
  \bibinfo{pages}{1501} (\bibinfo{year}{1998}).
%
\bibitem[{\citenamefont{Arndt et~al.}(2011)\citenamefont{Arndt, Stockert,
  Schmalzl, Faulhaber, Jeevan, Geibel, Schmidt, Loewenhaupt, and
  Steglich}}]{Arndt2011PRL}
\bibinfo{author}{\bibfnamefont{J.}~\bibnamefont{Arndt}},
  \bibinfo{author}{\bibfnamefont{O.}~\bibnamefont{Stockert}},
  \bibinfo{author}{\bibfnamefont{K.}~\bibnamefont{Schmalzl}},
  \bibinfo{author}{\bibfnamefont{E.}~\bibnamefont{Faulhaber}},
  \bibinfo{author}{\bibfnamefont{H.~S.} \bibnamefont{Jeevan}},
  \bibinfo{author}{\bibfnamefont{C.}~\bibnamefont{Geibel}},
  \bibinfo{author}{\bibfnamefont{W.}~\bibnamefont{Schmidt}},
  \bibinfo{author}{\bibfnamefont{M.}~\bibnamefont{Loewenhaupt}},
  \bibnamefont{and} \bibinfo{author}{\bibfnamefont{F.}~\bibnamefont{Steglich}},
  \bibinfo{journal}{Phys. Rev. Lett.} \textbf{\bibinfo{volume}{106}},
  \bibinfo{pages}{246401} (\bibinfo{year}{2011}).
%
\bibitem[{\citenamefont{Bredl et~al.}(1983)\citenamefont{Bredl, Spille,
  Rauchschwalbe, Lieke, Steglich, Cordier, Assmus, Herrmann, and
  Aarts}}]{Bredl1983JMMM}
\bibinfo{author}{\bibfnamefont{C.~D.} \bibnamefont{Bredl}},
  \bibinfo{author}{\bibfnamefont{H.}~\bibnamefont{Spille}},
  \bibinfo{author}{\bibfnamefont{U.}~\bibnamefont{Rauchschwalbe}},
  \bibinfo{author}{\bibfnamefont{W.}~\bibnamefont{Lieke}},
  \bibinfo{author}{\bibfnamefont{F.}~\bibnamefont{Steglich}},
  \bibinfo{author}{\bibfnamefont{G.}~\bibnamefont{Cordier}},
  \bibinfo{author}{\bibfnamefont{W.}~\bibnamefont{Assmus}},
  \bibinfo{author}{\bibfnamefont{M.}~\bibnamefont{Herrmann}}, \bibnamefont{and}
  \bibinfo{author}{\bibfnamefont{J.}~\bibnamefont{Aarts}}, \bibinfo{journal}{J.
  Magn. Magn. Mater.} \textbf{\bibinfo{volume}{31-34}}, \bibinfo{pages}{373}
  (\bibinfo{year}{1983}).
%
\bibitem[{\citenamefont{K\"{u}bert and Hirschfeld}(1998)}]{Kubert1998SSC}
\bibinfo{author}{\bibfnamefont{C.}~\bibnamefont{K\"{u}bert}} \bibnamefont{and}
  \bibinfo{author}{\bibfnamefont{P.~J.} \bibnamefont{Hirschfeld}},
  \bibinfo{journal}{Solid State Commun.} \textbf{\bibinfo{volume}{105}},
  \bibinfo{pages}{459} (\bibinfo{year}{1998}).
%
\bibitem[{\citenamefont{Bouquet et~al.}(2001)\citenamefont{Bouquet, Wang,
  Fisher, Hinks, Jorgensen, Junod, and Phillips}}]{Bouquet2001EPL}
\bibinfo{author}{\bibfnamefont{F.}~\bibnamefont{Bouquet}},
  \bibinfo{author}{\bibfnamefont{Y.}~\bibnamefont{Wang}},
  \bibinfo{author}{\bibfnamefont{R.~A.} \bibnamefont{Fisher}},
  \bibinfo{author}{\bibfnamefont{D.~G.} \bibnamefont{Hinks}},
  \bibinfo{author}{\bibfnamefont{J.~D.} \bibnamefont{Jorgensen}},
  \bibinfo{author}{\bibfnamefont{A.}~\bibnamefont{Junod}}, \bibnamefont{and}
  \bibinfo{author}{\bibfnamefont{N.~E.} \bibnamefont{Phillips}},
  \bibinfo{journal}{EuroPhys. Lett.} \textbf{\bibinfo{volume}{56}},
  \bibinfo{pages}{856} (\bibinfo{year}{2001}).

\bibitem[{\citenamefont{Yashima et~al.}(2009)\citenamefont{Yashima, Nishimura,
  Mukuda, Kitaoka, Miyazawa, Shirage, Kihou, Kito, Eisaki, and
  Iyo}}]{Yashima2009JPSJ}
\bibinfo{author}{\bibfnamefont{M.}~\bibnamefont{Yashima}},
  \bibinfo{author}{\bibfnamefont{H.}~\bibnamefont{Nishimura}},
  \bibinfo{author}{\bibfnamefont{H.}~\bibnamefont{Mukuda}},
  \bibinfo{author}{\bibfnamefont{Y.}~\bibnamefont{Kitaoka}},
  \bibinfo{author}{\bibfnamefont{K.}~\bibnamefont{Miyazawa}},
  \bibinfo{author}{\bibfnamefont{P.~M.} \bibnamefont{Shirage}},
  \bibinfo{author}{\bibfnamefont{K.}~\bibnamefont{Kihou}},
  \bibinfo{author}{\bibfnamefont{H.}~\bibnamefont{Kito}},
  \bibinfo{author}{\bibfnamefont{H.}~\bibnamefont{Eisaki}}, \bibnamefont{and}
  \bibinfo{author}{\bibfnamefont{A.}~\bibnamefont{Iyo}}, \bibinfo{journal}{J.
  Phys. Soc. Jpn.} \textbf{\bibinfo{volume}{78}}, \bibinfo{pages}{103702}
  (\bibinfo{year}{2009}).

\bibitem[{\citenamefont{Volovik}(1993)}]{Volovik1993JETPL}
\bibinfo{author}{\bibfnamefont{G.~E.} \bibnamefont{Volovik}},
  \bibinfo{journal}{JETP Lett.} \textbf{\bibinfo{volume}{58}},
  \bibinfo{pages}{469} (\bibinfo{year}{1993}).
%
\bibitem[{\citenamefont{Rauchschwalbe et~al.}(1982)\citenamefont{Rauchschwalbe,
  Lieke, Bredl, Steglich, Aarts, Martini, and Mota}}]{Rauchschwalbe1982PRL}
\bibinfo{author}{\bibfnamefont{U.}~\bibnamefont{Rauchschwalbe}},
  \bibinfo{author}{\bibfnamefont{W.}~\bibnamefont{Lieke}},
  \bibinfo{author}{\bibfnamefont{C.~D.} \bibnamefont{Bredl}},
  \bibinfo{author}{\bibfnamefont{F.}~\bibnamefont{Steglich}},
  \bibinfo{author}{\bibfnamefont{J.}~\bibnamefont{Aarts}},
  \bibinfo{author}{\bibfnamefont{K.~M.} \bibnamefont{Martini}},
  \bibnamefont{and} \bibinfo{author}{\bibfnamefont{A.~C.} \bibnamefont{Mota}},
  \bibinfo{journal}{Phys. Rev. Lett.} \textbf{\bibinfo{volume}{49}},
  \bibinfo{pages}{1448} (\bibinfo{year}{1982}).
%
\bibitem[{\citenamefont{Nakai et~al.}(2002)\citenamefont{Nakai, Ichioka, and
  Machida}}]{Nakai2002JPSJ}
\bibinfo{author}{\bibfnamefont{N.}~\bibnamefont{Nakai}},
  \bibinfo{author}{\bibfnamefont{M.}~\bibnamefont{Ichioka}}, \bibnamefont{and}
  \bibinfo{author}{\bibfnamefont{K.}~\bibnamefont{Machida}},
  \bibinfo{journal}{J. Phys. Soc. Jpn.} \textbf{\bibinfo{volume}{71}},
  \bibinfo{pages}{23} (\bibinfo{year}{2002}).
%
\bibitem[{\citenamefont{Nakai et~al.}(2004)\citenamefont{Nakai,
  Miranovi{$\acute{\mathrm{c}}$}, Ichioka, and
  Machida}}]{Nakai2004PRB}
\bibinfo{author}{\bibfnamefont{N.}~\bibnamefont{Nakai}},
  \bibinfo{author}{\bibfnamefont{P.}~\bibnamefont{Miranovi{$\acute{\mathrm{c}}$}}}, 
  \bibinfo{author}{\bibfnamefont{M.}~\bibnamefont{Ichioka}},
  \bibnamefont{and} \bibinfo{author}{\bibfnamefont{K.}~\bibnamefont{Machida}},
  \bibinfo{journal}{Phys. Rev. B} \textbf{\bibinfo{volume}{70}},
  \bibinfo{pages}{100503} (\bibinfo{year}{2004}).
%
\bibitem[{\citenamefont{Vekhter et~al.}(1999)\citenamefont{Vekhter, Hirschfeld,
  Carbotte, and Nicol}}]{Vekhter1999PRB}
\bibinfo{author}{\bibfnamefont{I.}~\bibnamefont{Vekhter}},
  \bibinfo{author}{\bibfnamefont{P.~J.} \bibnamefont{Hirschfeld}},
  \bibinfo{author}{\bibfnamefont{J.~P.} \bibnamefont{Carbotte}},
  \bibnamefont{and} \bibinfo{author}{\bibfnamefont{E.~J.} \bibnamefont{Nicol}},
  \bibinfo{journal}{Phys. Rev. B} \textbf{\bibinfo{volume}{59}},
  \bibinfo{pages}{R9023} (\bibinfo{year}{1999}).
%
\bibitem[{\citenamefont{Sakakibara et~al.}(2007)\citenamefont{Sakakibara,
  Yamada, Custers, Yano, Tayama, Aoki, and Machida}}]{Sakakibara2007JPSJ}
\bibinfo{author}{\bibfnamefont{T.}~\bibnamefont{Sakakibara}},
  \bibinfo{author}{\bibfnamefont{A.}~\bibnamefont{Yamada}},
  \bibinfo{author}{\bibfnamefont{J.}~\bibnamefont{Custers}},
  \bibinfo{author}{\bibfnamefont{K.}~\bibnamefont{Yano}},
  \bibinfo{author}{\bibfnamefont{T.}~\bibnamefont{Tayama}},
  \bibinfo{author}{\bibfnamefont{H.}~\bibnamefont{Aoki}}, \bibnamefont{and}
  \bibinfo{author}{\bibfnamefont{K.}~\bibnamefont{Machida}},
  \bibinfo{journal}{J. Phys. Soc. Jpn.} \textbf{\bibinfo{volume}{76}},
  \bibinfo{pages}{051004} (\bibinfo{year}{2007}).
%
\bibitem[{\citenamefont{Kittaka et~al.}(2012)\citenamefont{Kittaka, Aoki,
  Sakakibara, Sakai, Nakatsuji, Tsutsumi, Ichioka, and
  Machida}}]{Kittaka2012PRB}
\bibinfo{author}{\bibfnamefont{S.}~\bibnamefont{Kittaka}},
  \bibinfo{author}{\bibfnamefont{Y.}~\bibnamefont{Aoki}},
  \bibinfo{author}{\bibfnamefont{T.}~\bibnamefont{Sakakibara}},
  \bibinfo{author}{\bibfnamefont{A.}~\bibnamefont{Sakai}},
  \bibinfo{author}{\bibfnamefont{S.}~\bibnamefont{Nakatsuji}},
  \bibinfo{author}{\bibfnamefont{Y.}~\bibnamefont{Tsutsumi}},
  \bibinfo{author}{\bibfnamefont{M.}~\bibnamefont{Ichioka}}, \bibnamefont{and}
  \bibinfo{author}{\bibfnamefont{K.}~\bibnamefont{Machida}},
  \bibinfo{journal}{Phys. Rev. B} \textbf{\bibinfo{volume}{85}},
  \bibinfo{pages}{060505} (\bibinfo{year}{2012}).
%
\bibitem[{\citenamefont{Zwicknagl and Pulst}(1993)}]{Zwicknagl1993Physica}
\bibinfo{author}{\bibfnamefont{G.}~\bibnamefont{Zwicknagl}} \bibnamefont{and}
  \bibinfo{author}{\bibfnamefont{U.}~\bibnamefont{Pulst}},
  \bibinfo{journal}{Physica B \label{Zwicknagl1993Physica}}
  \textbf{\bibinfo{volume}{186-188}}, \bibinfo{pages}{895}
  (\bibinfo{year}{1993}).
%
\bibitem[{\citenamefont{Kuroki et~al.}(2008)\citenamefont{Kuroki, Onari, Arita,
  Usui, Tanaka, Kontani, and Aoki}}]{Kuroki2008PRL}
\bibinfo{author}{\bibfnamefont{K.}~\bibnamefont{Kuroki}},
  \bibinfo{author}{\bibfnamefont{S.}~\bibnamefont{Onari}},
  \bibinfo{author}{\bibfnamefont{R.}~\bibnamefont{Arita}},
  \bibinfo{author}{\bibfnamefont{H.}~\bibnamefont{Usui}},
  \bibinfo{author}{\bibfnamefont{Y.}~\bibnamefont{Tanaka}},
  \bibinfo{author}{\bibfnamefont{H.}~\bibnamefont{Kontani}}, \bibnamefont{and}
  \bibinfo{author}{\bibfnamefont{H.}~\bibnamefont{Aoki}},
  \bibinfo{journal}{Phys. Rev. Lett.} \textbf{\bibinfo{volume}{101}},
  \bibinfo{pages}{087004} (\bibinfo{year}{2008}).
%
\bibitem[{\citenamefont{Ichioka and Machida}(2007)}]{Ichioka2007PRB}
\bibinfo{author}{\bibfnamefont{M.}~\bibnamefont{Ichioka}} \bibnamefont{and}
  \bibinfo{author}{\bibfnamefont{K.}~\bibnamefont{Machida}},
  \bibinfo{journal}{Phys. Rev. B} \textbf{\bibinfo{volume}{76}},
  \bibinfo{pages}{064502} (\bibinfo{year}{2007}).
%
\bibitem[{\citenamefont{Machida and Ichioka}(2008)}]{Machida2008PRB}
\bibinfo{author}{\bibfnamefont{K.}~\bibnamefont{Machida}} \bibnamefont{and}
  \bibinfo{author}{\bibfnamefont{M.}~\bibnamefont{Ichioka}},
  \bibinfo{journal}{Phys. Rev. B} \textbf{\bibinfo{volume}{77}},
  \bibinfo{pages}{184515} (\bibinfo{year}{2008}).
%
\bibitem[{\citenamefont{Matsuda and Shimahara}(2007)}]{Matsuda2007JPSJ}
\bibinfo{author}{\bibfnamefont{Y.}~\bibnamefont{Matsuda}} \bibnamefont{and}
  \bibinfo{author}{\bibfnamefont{H.}~\bibnamefont{Shimahara}},
  \bibinfo{journal}{J. Phys. Soc. Jpn.} \textbf{\bibinfo{volume}{76}},
  \bibinfo{pages}{051005} (\bibinfo{year}{2007}), \bibinfo{note}{and references
  therein}.
%
\bibitem[{\citenamefont{Nakai et~al.}(2006)\citenamefont{Nakai,
  Miranovi{$\acute{\mathrm{c}}$}, Ichioka, and
  Machida}}]{Nakai2006PRB}
\bibinfo{author}{\bibfnamefont{N.}~\bibnamefont{Nakai}},
  \bibinfo{author}{\bibfnamefont{P.}~\bibnamefont{Miranovi{$\acute{\mathrm{c}}$}}}, 
  \bibinfo{author}{\bibfnamefont{M.}~\bibnamefont{Ichioka}},
  \bibnamefont{and} \bibinfo{author}{\bibfnamefont{K.}~\bibnamefont{Machida}},
  \bibinfo{journal}{Phys. Rev. B} \textbf{\bibinfo{volume}{73}},
  \bibinfo{pages}{172501} (\bibinfo{year}{2006}).

\bibitem[{\citenamefont{Machida et~al.}(2013)\citenamefont{Machida}}]{Machida}
\bibinfo{author}{\bibfnamefont{Y.}~\bibnamefont{Tsutsumi}},
  \bibinfo{author}{\bibfnamefont{M.}~\bibnamefont{Ichioka}},
  \bibnamefont{and} \bibinfo{author}{\bibfnamefont{K.}~\bibnamefont{Machida}},
  \bibinfo{memo}{in preparation}.

\end{thebibliography}

\begin{thebibliography}{13}
\expandafter\ifx\csname natexlab\endcsname\relax\def\natexlab#1{#1}\fi
\expandafter\ifx\csname bibnamefont\endcsname\relax
  \def\bibnamefont#1{#1}\fi
\expandafter\ifx\csname bibfnamefont\endcsname\relax
  \def\bibfnamefont#1{#1}\fi
\expandafter\ifx\csname citenamefont\endcsname\relax
  \def\citenamefont#1{#1}\fi
\expandafter\ifx\csname url\endcsname\relax
  \def\url#1{\texttt{#1}}\fi
\expandafter\ifx\csname urlprefix\endcsname\relax\def\urlprefix{URL }\fi
\providecommand{\bibinfo}[2]{#2}
\providecommand{\eprint}[2][]{\url{#2}}

\bibitem[{\citenamefont{Blaha et~al.}()\citenamefont{Blaha, Schwarz, Madsen,
  Kvasnicka, and Luitz}}]{rf:Blaha}
\bibinfo{author}{\bibfnamefont{P.}~\bibnamefont{Blaha}},
  \bibinfo{author}{\bibfnamefont{K.}~\bibnamefont{Schwarz}},
  \bibinfo{author}{\bibfnamefont{G.~K.~H.} \bibnamefont{Madsen}},
  \bibinfo{author}{\bibfnamefont{D.}~\bibnamefont{Kvasnicka}},
  \bibnamefont{and} \bibinfo{author}{\bibfnamefont{J.}~\bibnamefont{Luitz}}, 
  \bibinfo{note}{WIEN2K (Karlheinz Schwarz, Techn. Universitat, Wien, Austria, 2001)}.

\bibitem[{\citenamefont{Perdew et~al.}(1996)\citenamefont{Perdew, Burke, and
  Ernzerhof}}]{rf:Perdew}
\bibinfo{author}{\bibfnamefont{J.~P.} \bibnamefont{Perdew}},
  \bibinfo{author}{\bibfnamefont{K.}~\bibnamefont{Burke}}, \bibnamefont{and}
  \bibinfo{author}{\bibfnamefont{M.}~\bibnamefont{Ernzerhof}},
  \bibinfo{journal}{Phys. Rev. Lett.} \textbf{\bibinfo{volume}{77}},
  \bibinfo{pages}{3865} (\bibinfo{year}{1996}).

\bibitem[{\citenamefont{Jarlborg et~al.}(1983)\citenamefont{Jarlborg, Braun,
  and Peter}}]{rf:Jarlborg}
\bibinfo{author}{\bibfnamefont{T.}~\bibnamefont{Jarlborg}},
  \bibinfo{author}{\bibfnamefont{H.~F.} \bibnamefont{Braun}}, \bibnamefont{and}
  \bibinfo{author}{\bibfnamefont{M.}~\bibnamefont{Peter}}, \bibinfo{journal}{Z.
  Phys. B} \textbf{\bibinfo{volume}{52}}, \bibinfo{pages}{295}
  (\bibinfo{year}{1983}).

\bibitem[{\citenamefont{Harima and Yanase}(1991)}]{rf:Harima}
\bibinfo{author}{\bibfnamefont{H.}~\bibnamefont{Harima}} \bibnamefont{and}
  \bibinfo{author}{\bibfnamefont{A.}~\bibnamefont{Yanase}},
  \bibinfo{journal}{J. Phys. Soc. Jpn.} \textbf{\bibinfo{volume}{60}},
  \bibinfo{pages}{21} (\bibinfo{year}{1991}).

\bibitem[{\citenamefont{Marzari and Vanderbilt}(1997)}]{rf:Marzari}
\bibinfo{author}{\bibfnamefont{N.}~\bibnamefont{Marzari}} \bibnamefont{and}
  \bibinfo{author}{\bibfnamefont{D.}~\bibnamefont{Vanderbilt}},
  \bibinfo{journal}{Phys. Rev. B} \textbf{\bibinfo{volume}{56}},
  \bibinfo{pages}{12847} (\bibinfo{year}{1997}).

\bibitem[{\citenamefont{Souza et~al.}(2001)\citenamefont{Souza, Marzari, and
  Vanderbilt}}]{rf:Souza}
\bibinfo{author}{\bibfnamefont{I.}~\bibnamefont{Souza}},
  \bibinfo{author}{\bibfnamefont{N.}~\bibnamefont{Marzari}}, \bibnamefont{and}
  \bibinfo{author}{\bibfnamefont{D.}~\bibnamefont{Vanderbilt}},
  \bibinfo{journal}{Phys. Rev. B} \textbf{\bibinfo{volume}{65}},
  \bibinfo{pages}{035109} (\bibinfo{year}{2001}).

\bibitem[{\citenamefont{Mostofi et~al.}(2008)\citenamefont{Mostofi, Yates, Lee,
  Souza, Vanderbilt, and Marzari}}]{rf:Mostofi}
\bibinfo{author}{\bibfnamefont{A.~A.} \bibnamefont{Mostofi}},
  \bibinfo{author}{\bibfnamefont{J.~R.} \bibnamefont{Yates}},
  \bibinfo{author}{\bibfnamefont{Y.~-S.} \bibnamefont{Lee}},
  \bibinfo{author}{\bibfnamefont{I.}~\bibnamefont{Souza}},
  \bibinfo{author}{\bibfnamefont{D.}~\bibnamefont{Vanderbilt}},
  \bibnamefont{and} \bibinfo{author}{\bibfnamefont{N.}~\bibnamefont{Marzari}},
  \bibinfo{journal}{Comput. Phys. Commun.} \textbf{\bibinfo{volume}{178}},
  \bibinfo{pages}{685} (\bibinfo{year}{2008}),
  \bibinfo{note}{{http://www.wannier.org/}}.

\bibitem[{\citenamefont{{J. Kune\v{s} {\it et al.}}}(2010)}]{rf:Kunes}
\bibinfo{author}{\bibnamefont{{J. Kune\v{s} {\it et al.}}}},
  \bibinfo{journal}{Comput. Phys. Commun.} \textbf{\bibinfo{volume}{181}},
  \bibinfo{pages}{1888} (\bibinfo{year}{2010}),
  \bibinfo{note}{{http://www.wien2k.at/reg\_user/unsupported/wien2wannier}}.

\bibitem[{\citenamefont{Anisimov et~al.}(1997)\citenamefont{Anisimov,
  Aryasetiawa, and Lichtenstein}}]{rf:Anisimov}
\bibinfo{author}{\bibfnamefont{V.~I.} \bibnamefont{Anisimov}},
  \bibinfo{author}{\bibfnamefont{F.}~\bibnamefont{Aryasetiawa}},
  \bibnamefont{and} \bibinfo{author}{\bibfnamefont{A.~I.}
  \bibnamefont{Lichtenstein}}, \bibinfo{journal}{J. Phys.: Condens. Matter}
  \textbf{\bibinfo{volume}{9}}, \bibinfo{pages}{767} (\bibinfo{year}{1997}).

\bibitem[{\citenamefont{Suzuki and Harima}(2010)}]{rf:Suzuki}
\bibinfo{author}{\bibfnamefont{M.~-T.} \bibnamefont{Suzuki}} \bibnamefont{and}
  \bibinfo{author}{\bibfnamefont{H.}~\bibnamefont{Harima}},
  \bibinfo{journal}{J. Phys. Soc. Jpn.} \textbf{\bibinfo{volume}{79}},
  \bibinfo{pages}{024705} (\bibinfo{year}{2010}).

\bibitem[{\citenamefont{Kang et~al.}(1990)\citenamefont{Kang, Allen,
  Gunnarsson, Christensen, Andersen, Lassailly, Maple, and
  Torikachvili}}]{rf:Kang}
\bibinfo{author}{\bibfnamefont{J.~-S.} \bibnamefont{Kang}},
  \bibinfo{author}{\bibfnamefont{J.~W.} \bibnamefont{Allen}},
  \bibinfo{author}{\bibfnamefont{O.}~\bibnamefont{Gunnarsson}},
  \bibinfo{author}{\bibfnamefont{N.~E.} \bibnamefont{Christensen}},
  \bibinfo{author}{\bibfnamefont{O.~K.} \bibnamefont{Andersen}},
  \bibinfo{author}{\bibfnamefont{Y.}~\bibnamefont{Lassailly}},
  \bibinfo{author}{\bibfnamefont{M.~B.} \bibnamefont{Maple}}, \bibnamefont{and}
  \bibinfo{author}{\bibfnamefont{M.~S.} \bibnamefont{Torikachvili}},
  \bibinfo{journal}{Phys. Rev. B} \textbf{\bibinfo{volume}{41}},
  \bibinfo{pages}{6610} (\bibinfo{year}{1990}).

\bibitem[{\citenamefont{Zwicknagl and Pulst}(1993)}]{rf:Zwicknagl}
\bibinfo{author}{\bibfnamefont{G.}~\bibnamefont{Zwicknagl}} \bibnamefont{and}
  \bibinfo{author}{\bibfnamefont{U.}~\bibnamefont{Pulst}},
  \bibinfo{journal}{Physica B} \textbf{\bibinfo{volume}{186-188}},
  \bibinfo{pages}{895} (\bibinfo{year}{1993}).

\bibitem[{\citenamefont{Eremin et~al.}(2008)\citenamefont{Eremin, Zwicknagl,
  Thalmeier, and Fulde}}]{rf:Eremin}
\bibinfo{author}{\bibfnamefont{I.}~\bibnamefont{Eremin}},
  \bibinfo{author}{\bibfnamefont{G.}~\bibnamefont{Zwicknagl}},
  \bibinfo{author}{\bibfnamefont{P.}~\bibnamefont{Thalmeier}},
  \bibnamefont{and} \bibinfo{author}{\bibfnamefont{P.}~\bibnamefont{Fulde}},
  \bibinfo{journal}{Phys. Rev. Lett. \label{Eremin}}
  \textbf{\bibinfo{volume}{101}}, \bibinfo{pages}{187001}
  (\bibinfo{year}{2008}).

\end{thebibliography}

\end{document}